\documentclass[twocolumn,floatfix,amssymb,amsmath,secnumarabic,nofootinbib]{revtex4-1}    
\usepackage[pdftex]{graphicx} 
\usepackage{dcolumn}  
\usepackage{bm}       
\usepackage[usenames,dvipsnames]{color}
\definecolor{URLCOL}{rgb}{0,0.52,0.83} 
\definecolor{LINKCOL}{rgb}{0.05,0.5,0} 
\definecolor{CITECOL}{rgb}{0.25,0,0.48} 
\usepackage{epstopdf}
\usepackage[pdftex,bookmarks,breaklinks,bookmarksopen,bookmarksnumbered,colorlinks,linkcolor=LINKCOL,linktocpage,citecolor=CITECOL,urlcolor=URLCOL,pdfpagemode=UseOutline,pdftex]{hyperref}

\def\preprintlink{ \href{http://dft.uci.edu}{dft.uci.edu} }
\def\preprinttext{PREPRINT}

\usepackage{fancyhdr}
\makeatletter
\fancypagestyle{titlepage}
{
	\lhead{\textsc{\preprinttext\ from the \href{http://dft.uci.edu/publications.php}{Burke Group archives}}}
	\chead{}
	\rhead{\preprintlink}
	\lfoot{}
}
\makeatother

\usepackage{amsmath, amsthm, amssymb}
\usepackage{graphicx}
\usepackage{natbib}

\renewcommand\Im{\operatorname{Im}}

\newcommand{\ve}{\varepsilon}

\begin{document}

\title{Kondo effect given exactly by density functional theory}

\author{Justin P. Bergfield}

\author{Zhen-Fei Liu}

\author{Kieron Burke}
\affiliation{Departments of Chemistry and Physics, University of California, Irvine, California 92697, USA}
\author{Charles A. Stafford}
\affiliation{Department of Physics, University of Arizona, 1118 East Fourth Street, Tucson, AZ 85721}

\date{\today}

\begin{abstract}
Transport through an {\em Anderson junction} (two
macroscopic electrodes coupled to an Anderson impurity)
is dominated by a Kondo peak 
in the spectral function at zero temperature.
The exact
single-particle Kohn-Sham potential
of density functional theory reproduces the linear transport exactly,
despite the lack of a Kondo peak in its spectral function.
Using Bethe ansatz techniques, we calculate this potential exactly for all coupling strengths, including
the cross-over from mean-field behavior to charge quantization caused by the derivative discontinuity.
A simple and accurate interpolation formula is also given.
\end{abstract}

\maketitle

It is a truth universally acknowledged that many-body effects in strongly correlated
systems are not reproduced by mean-field theory.
Although Kohn-Sham (KS) density functional theory (DFT) is formally exact, it is
a non-interacting theory yielding only the ground-state
energy and density of a system.  No other information about the
correlated many-body wavefunction is available.
Dynamical properties, such as excitations and response functions,  are
also not predicted by ground-state DFT, even with the exact functional \cite{DG90}.  
The hope is that, for weakly-correlated systems in which ground-state DFT approximations perform
well for total energies, geometries, etc., the errors in such calculations are small.
Nothing in the 
theorems of DFT guarantees that a ground-state KS calculation can describe
transport correctly \cite{Koentopp08}.

\begin{figure}[tb]
	\centering	
	\includegraphics{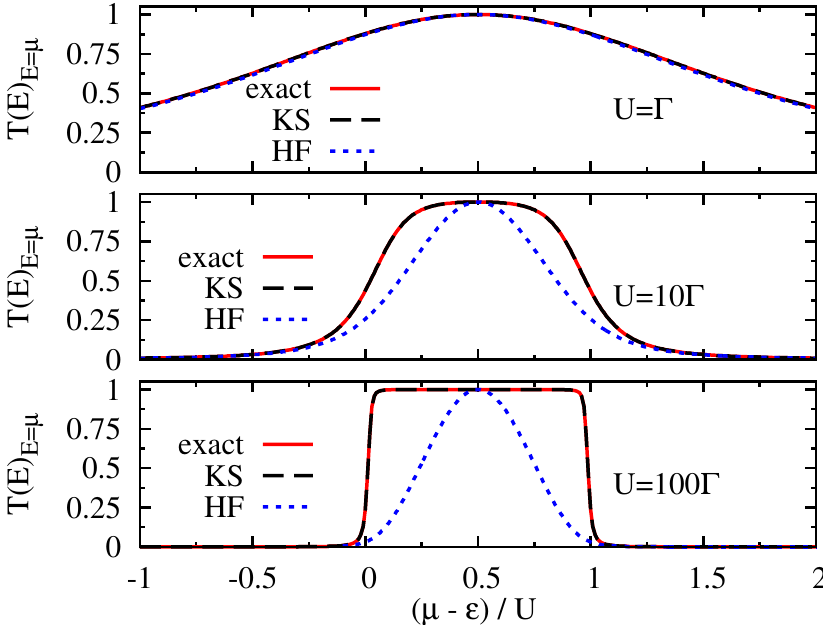}
	\vspace{-.25cm}
\caption{(Color online) Zero-temperature transmission of an Anderson junction as a 
function of $\ve$  
using Bethe ansatz (exact), 
Kohn-Sham DFT with the exact functional (KS),
and (spin-restricted) Hartree-Fock (HF). As $U$ increases, HF misses the sharp structure,
but the KS transport is always exact! }
	\label{fig:k}
	\vspace{-.4cm}
\end{figure}

Consider transport through an
Anderson junction \cite{Anderson61,Glazman04},
composed of two macroscopic leads coupled to an Anderson impurity.
As an integrable system, the Anderson model is a paradigm of many-body physics.
It is also an accurate model of the low-energy spectrum
of molecular radical-based junctions \cite{Bergfield11}.  
In Fig.~\ref{fig:k},
 we show the zero-temperature transmission through an Anderson junction as a
function of the 
energy $\varepsilon$ of the resonant level 
using Bethe ansatz (BA), exact Kohn-Sham (KS) DFT, and Hartree-Fock (HF).  
In the figure, $\mu$ is the chemical potential (Fermi energy) of the metal electrodes.  
Remarkably, the KS-DFT treatment of this problem reproduces
the transmission {\em exactly}, apparently describing the non-perturbative Kondo
effect whose spectral peak 
is the source of the perfect transmission 
when $\varepsilon < \mu < \varepsilon+U$.  

The inability to describe sharp steps in transmission is a well-understood failure of
standard density functional approximations.  In the limit of weak
coupling to the leads, the system is a prototype example where
the effects of the infamous derivative discontinuity is seen \cite{PPLB82}.
For such a system, the exchange-correlation (XC) energy of the molecule is strictly linear
between integer values, and so the XC potential, its functional derivative,
jumps discontinuously at such values \cite{PPLB82}.  This effect has been implicated
in many well-known failures of DFT approximations such as strongly-correlated systems \cite{MCY09},  
charge-transfer excitations \cite{M05}, and over-estimation of
the current in organic junctions \cite{EWK04}.

In this letter, we (a) solve the Anderson junction using BA and
invert the KS equations to {\em derive} the exact KS potential,
(b) show that the transport is recovered exactly, but only for zero
temperature and weak bias, and (c) parametrize the XC potential, providing
a unique interpolation formula that works for all correlation strengths.
The exact and KS-DFT transport are intimately 
related via 
the Friedel sum-rule, a statement equating the occupancy and the
transmission phase at the Fermi energy, as has
been discussed for nearly isolated resonances \cite{Schmitteckert08,Mera10} or
single-mode leads \cite{Mera10}.  Since our system fits neither of these
categories, it reopens the question of just when an accurate ground-state
KS calculation will yield accurate transport.

Using BA \cite{Wiegmann83,Konik01},
we calculate the exact KS potential 
for the Anderson model, and investigate how the derivative discontinuity develops in the limit of weak
impurity-lead coupling.
From this solution, 
we extract the exact asymptotic scaling form of the derivative discontinuity and establish an interpolation formula for the KS potential 
which is accurate for all coupling strengths, computationally simple, and illustrates the crossover from mean-field behavior at strong impurity-lead coupling
to charge quantization at weak coupling.

Consider a junction 
composed of a nanoscale central region (C) connected to two macroscopic electrodes,
labeled left (L) and right (R).  
The Hamiltonian of the system is ${\cal H}={\cal H}_{\rm C}+{\cal H}_{\rm T}+{\cal H}_{\rm R}+{\cal H}_{\rm L}$, 
where ${\cal H}_{\rm R/L}$ describe Fermi gases in the electrodes and ${\cal H}_{\rm T}$ describes tunneling between the central region and the electrodes.
The central region has the Hamiltonian \cite{Anderson61}:
\begin{equation}
	{\cal H}_{\rm C} = \ve (\hat{n}_\uparrow + \hat{n}_\downarrow) + U n_\uparrow n_\downarrow,
	\label{eq:H_anderson}
\end{equation}
where $\hat{n}_\sigma = d_\sigma^\dagger d_\sigma$ is the number operator for
spin $\sigma$ and the charging energy $U$ is given by the Coulomb integral.  
The exact Green's function of an Anderson junction may be found using Dyson's equation in an orthonormal basis:
\begin{equation}
	G(E)= \left[E- \ve - \Sigma_{\rm C}(E) - \Sigma_{\rm T}(E)\right]^{-1},
	\label{eq:G_many_body}
\end{equation}
where $\Sigma_{\rm C}$ is the Coulomb self-energy 
and $\Sigma_{\rm T}\equiv \Sigma_{\rm T}^{\rm R}+\Sigma_{\rm T}^{\rm L}$ is the tunneling self-energy \cite{Bergfield09}.
We consider transport in the broad-band limit where $\Sigma^\alpha_{\rm T}(E) = -i\Gamma^\alpha/2$ is pure imaginary and independent of energy, 
and define the mean tunneling-width $\Gamma = (\Gamma^L + \Gamma^R)/2$. 

\begin{figure}[tb]
	\centering	\includegraphics{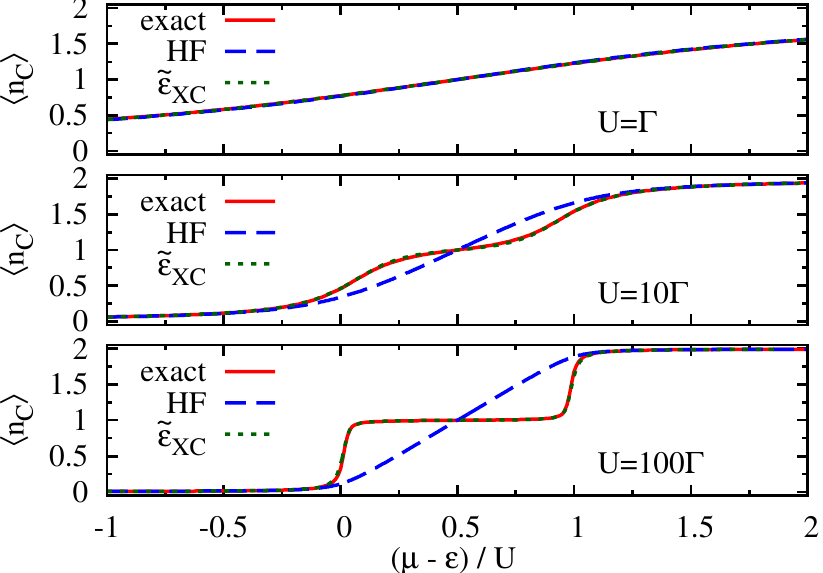}
		\vspace{-.25cm}
\caption{(Color online) The occupancy of the Anderson junction as a function of $\mu-\ve$.  
The BA results (solid line) exhibit discrete charge steps when $U \gg \Gamma$, whereas the mean-field RHF (dashed line) never does.  The approximate occupancies calculated self-consistently within the KS-scheme using $\tilde{\ve}_{XC}$ (Eq.~(\ref{eq:approx_eps_xc})) (dotted line) are nearly indistinguishable from the exact BA results.  The exact occupancies were used to generate Fig.~\ref{fig:k}.}
	\label{fig:occupany}
	\vspace{-.4cm}
\end{figure}

At zero temperature,
the linear-response transmission function of an Anderson junction is given by \cite{Glazman04,Konik01} 
\begin{equation}
	T(E)|_{E=\mu} = \Gamma^{\rm L}\Gamma^{\rm R}|G(\mu)|^2=
	\frac{4\Gamma^{\rm L}\Gamma^{\rm R}}{\left(\Gamma^{\rm L}+\Gamma^{\rm R}\right)^2}  \sin^2\left[\frac{\theta(\mu)}{2} \right],
	\label{eq:Tfrmn}
\end{equation}
where  
$\theta(\mu)$ is the sum of transmission eigenphases 
evaluated at the Fermi energy $\mu = \mu_L = \mu_R$.  The total number of  
electrons on the central impurity is related to $\theta(\mu)$ at zero temperature by the Friedel sum-rule \cite{Langreth66, Friedel58,MKNS10} 
\begin{equation}
	\langle n_{\rm C} \rangle =  \theta\left(\mu\right)/{\pi}.
	\label{eq:friedel}
\end{equation}
From the Bethe ansatz, one finds \cite{Wiegmann83} 
\begin{equation}
\begin{split}
\langle n_{\rm C}(\mu) \rangle = 1   -\frac{i}{\pi \sqrt{2}}   \int_{-\infty}^\infty & \frac{d\omega}{\omega+ i\eta} \frac{e^{-|\omega|/2}}{G^{(-)}(\omega)} \\
&  \int_{-\infty}^{\infty} e^{i\omega( g(k)-Q)}\Delta(k) dk
 \label{eq:ba1}
\end{split}
\end{equation}
with Q defined by the condition
\begin{equation}
\!\frac{U - 2\mu}{\sqrt{2U\Gamma}} = \frac{i}{\sqrt{2\pi}} \int_{-\infty}^\infty \frac{d\omega}{\omega + i\eta} \frac{e^{-|\omega|/2 - i\omega Q}}{G^{(-)}(\omega)} \frac{1}{\sqrt{-i\omega+\eta}},
\end{equation}
making use of the following functions
\begin{equation}
	G^{(-)}(\omega) = \frac{\sqrt{2\pi}}{ {\bf \Gamma}(1/2 + i\omega/2\pi)} \left( \frac{i\omega+\eta}{2\pi e}\right)^{i\omega/2\pi},
	\label{eq:ba3}
\end{equation}
$g(k) = (k+\mu - U/2)^2 / (2U\Gamma)$, and $\Delta(k) = (1/\pi) \Gamma/(\Gamma^2 + (k+\mu)^2)$,
where ${\bf \Gamma}(x)$ is the Gamma function and
$\eta=0^+$.    
$\langle n_{\rm C} \rangle$ is plotted as a function of $\ve$ in Fig.~\ref{fig:occupany}.

The KS ansatz of DFT employs an effective single-particle
description, defined to reproduce the ground-state density of the interacting system.  
By the Hohenberg-Kohn theorem, this potential is unique if it exists (it usually does) \cite{DG90}.
The relationship between potential and density is fixed only in the full basis-set
limit, but can be defined for lattice models.  
The leads of an Anderson model are non-interacting and characterized by a total charge, which remains constant.  Thus the definition of the KS system in this extreme case is that of a
non-interacting junction ($U=0$) with an on-site potential chosen to reproduce the
on-site occupancy of the interacting system.

The KS Green's function in the central region may be written as 
\begin{equation}
	G^{\rm s}(E) = \left(E-\ve^{\rm s} + i \Gamma \right)^{-1},
\label{eq:KS_Greens_fcn}
\end{equation}
where $\ve^{\rm s}$ is the KS potential for an 
electron
on the impurity, and is written
\begin{equation}
\ve^{\rm s}= \ve + U \langle n_{\rm C} \rangle /2 + \ve_{\rm XC},
\label{eq:KS_onsite}
\end{equation}
where the second term is the Hartree contribution and the last is 
the correlation potential (there are no exchange contributions).  
The Anderson model has no internal molecular structure so the KS lead-molecule coupling is $\Gamma$ \cite{Bergfield11,Evers11b}.  For more complex systems, $\Gamma$ need not be equal to the KS lead-molecule coupling.
In a standard DFT calculation, $\ve_{\rm XC}$ 
is approximated as a functional of the density\cite{DG90}.
The occupancy of the central region is 
\begin{equation}
	\langle n_{\rm C} \rangle =  2 \int_{-\infty}^{\infty} \frac{dE}{2\pi} \Im \left[ G(E) \right]^<,
	\label{eq:KS_n_mol}
\end{equation}
where
the ``lesser'' Green's function is found using the Keldysh 
relation \cite{Bergfield09}
\begin{equation}
	\left[G(E) \right]^< = 2if(E) \Gamma \left| G(E) \right|^2,
\end{equation}
where at zero temperature,
 $f(E)\equiv \Theta(\mu-E)$ and $\Theta$ is the Heaviside function.  
Inserting the
KS Green's function and solving Eq.~(\ref{eq:KS_n_mol}) for $\ve_{\rm s}$ gives
\begin{equation}
	\ve^{\rm s} = \mu + \Gamma \cot\left(\frac{\pi}{2}\langle n_{\rm C} \rangle  \right),
	\label{eq:vs}
\end{equation}
which defines the {\em exact} KS potential within the Anderson model. 
The KS transmission is then
\begin{equation}
	T^{\rm s}(E) = \frac{\Gamma^{\rm L} \Gamma^{\rm R}}{\left(E-\ve^{\rm s}\right)^2 +\Gamma^2}.
	\label{eq:KS_Transmission_fcn}
\end{equation}
Plugging Eq.~(\ref{eq:vs}) into Eq.~(\ref{eq:KS_Transmission_fcn}),
 we find that $T^{\rm s}(E)$ is identical to $T(E)$, 
as was shown in Fig.~\ref{fig:k}.  
Although this identity can be derived using, e.g., local Fermi liquid theory \cite{Glazman04}, nonetheless its significance is profound: If (and only if) a mean-field theory yields the exact occupation will it yield the exact transmission.

\begin{figure}[tb]
	\centering	
	\includegraphics{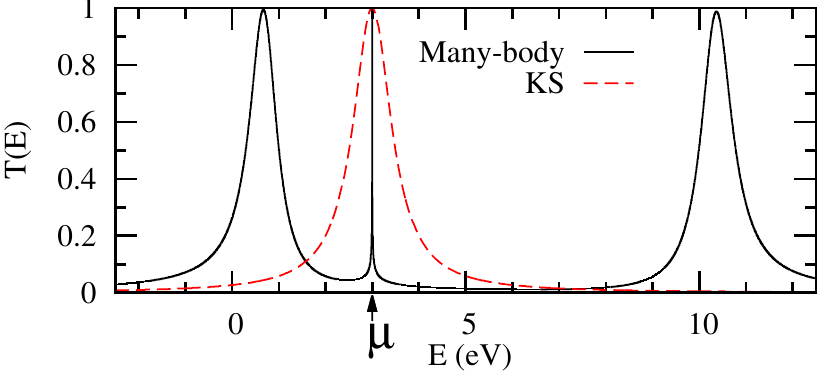}
	\vspace{-.25cm}
	\caption{(Color online) Transmission through an Anderson junction at
fixed $\mu=3eV$ with $\ve=0$, $\Gamma=0.5eV$ and $U =10eV$, so that  
$T_{\rm K}=2mK$ \cite{Haldane78,Wiegmann83,Konik01}.  
The Doniach-Sunjic form \cite{Doniach70} 
of the spectral function \cite{Silver90} is used near the Kondo peak, while 
the non-singular portion is calculated using 
the methods of Refs.~\cite{Bergfield11,Bergfield09} with the exact occupancy. 
Logarithmic shifts of the charging resonances are also included
\cite{Haldane78}.  
More sophisticated numerical methods \cite{Costi94} reproduce the qualitative features shown here.}
\label{fig:spectral_fcn_exact_vs_KS}
\vspace{-.4cm}
\end{figure}

In an Anderson junction, the Friedel sum-rule
connects the transmission at the Fermi energy to the occupancy at zero temperature.  
As shown in Fig.~\ref{fig:spectral_fcn_exact_vs_KS}, the full transmission
 spectrum of an Anderson junction exhibits three peaks: 
Two Coulomb-blockade peaks of width $\sim\Gamma$ centered about $E \approx 0$ and $E\approx U$ 
and a third zero-bias Kondo peak of width $\sim k_B T_K$  
pinned at $E=\mu$.
In contrast, the KS-DFT transmission spectrum is a single
Lorentzian of width $\Gamma$ peaked at $E=\mu$.  As indicated in the figure,
the KS value is a huge overestimate anywhere
more than several $k_B T_K$
away from $\mu$, implying that
the ground-state KS potential
does not accurately predict transport at temperatures larger
than $T_K$ or for bias voltages larger than $k_B T_{K}/e$.

\begin{figure*}[tb]
	\centering	
	\includegraphics{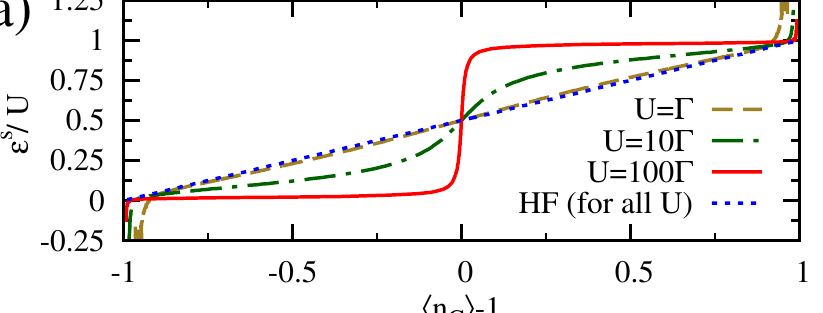}	
	\includegraphics{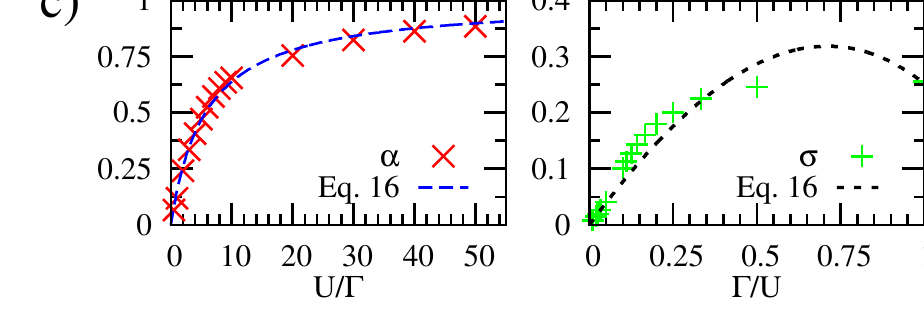}
	\includegraphics{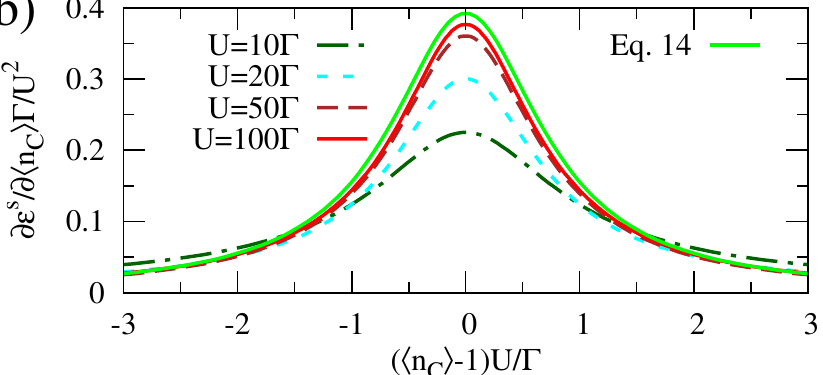}
\includegraphics{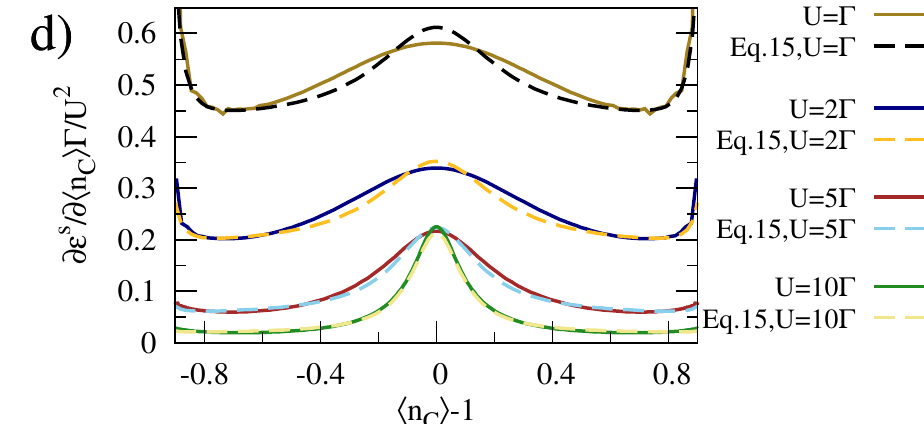}
	\vspace{-.6cm}
\caption{(Color online) The exact KS potential $\ve^{\rm s}$ of the Anderson junction.
(a) For $U \approx \Gamma$, HF is accurate; as $U$ grows, an increasingly
sharp step develops, due
to the {\em derivative discontinuity} of the exchange-correlation energy
as a function of particle number.
(b) $\partial \ve^{\rm s}/\partial \langle n_C \rangle$
for several large values of $U/\Gamma$.  
As $U/\Gamma \to \infty$, the exact (Lorentzian) asymptotic scaling 
form is recovered.  
(c) Exact and interpolated $\alpha$ and 
$\sigma$ functions are shown
in the left and right-hand panels, respectively, showing the  
crossover from the meanfield ($U/\Gamma\sim 0$) to the strongly correlated
regime ($U/\Gamma\rightarrow \infty$).  
The crossover from weakly to strongly correlated occurs for 
$U/\Gamma\approx 6$.  
(d) Exact and approximate 
$\partial \ve^{\rm s}/\partial \langle n_C \rangle$ spectra
for several moderate  $U/\Gamma$ values, highlighting the accuracy
of Eq.~(\ref{eq:approx_eps_xc}).  The dashed lines use
Eq.~(\ref{eq:s_ansatz}).  
	}
	\label{fig:deriv_discontinuity}
	\vspace{-.4cm}
\end{figure*}

From Eqs.~(\ref{eq:Tfrmn}), (\ref{eq:friedel}), (\ref{eq:vs}) and (\ref{eq:KS_Transmission_fcn}) 
it is evident that the HF errors in transmission
in Fig. ~\ref{fig:k} stem from errors in the occupancies of Fig. ~\ref{fig:occupany}.
When $U \lesssim \Gamma$, HF yields accurate occupancies and transmissions.
But for $U \gg \Gamma$,
the HF occupancies lack the distinct steps present in
the exact solutions, causing corresponding discrepancies in 
the  transport.
Qualitatively similar errors would be found with any local or semilocal
approximation for the XC potential, because such approximations are
smooth functions of the interaction strength \cite{KBE06}.  
But the exact KS potential of an isolated system, infinitely weakly coupled
to a reservoir, displays discontinuous jumps at integer particle number
\cite{PPLB82}, just as ours does as $\Gamma/U \rightarrow 0$.  The
KS potential $\ve^{\rm s}$ is shown as a function of occupancy
in Fig.\ \ref{fig:deriv_discontinuity}a for several values of 
$U/\Gamma$.  The HF potential is linear with a slope of $U/2$.
For large but finite $U/\Gamma$, the KS potential is not discontinuous but has
steps (of width $\sim\Gamma/U$) corresponding to the plateaus in the occupancy,
becoming discontinuous in the limit.
When $U/\Gamma$ is sufficiently small there is no step in
the KS potential and the 
HF approximation is accurate.    

We now show how the step
develops as $\Gamma/U\rightarrow 0$. 
In Fig.\ \ref{fig:deriv_discontinuity}a, $\ve^{\rm s}$ develops a step of height $U$ at $\langle n_{\rm C}\rangle=1$
whose width decreases as $\Gamma/U\rightarrow 0$.
Fig.\ \ref{fig:deriv_discontinuity}b shows
the exact derivative  
$\partial \ve^{\rm s}/ \partial \langle n_{\rm C} \rangle$  
in the vicinity of $\langle n_{\rm C} \rangle=1$.  
The horizontal and vertical axes are rescaled
to illustrate the scaling behavior of the step as $\Gamma/U\rightarrow 0$.
From the BA solution, as $U\to\infty$ \cite{tbp}:
\begin{equation}
{\ve}_{\rm XC} \simeq \frac{U}{2}\left[1-\langle n_{\rm C} \rangle -
\frac{2}{\pi}\tan^{-1}\left[\frac{\pi^2U(1-\langle n_{\rm C} \rangle)}
{8\Gamma}\right] \right],
\label{eq:eps_xc_inf} 
\end{equation}
whose derivative yields a Lorentzian.
In Fig.\ \ref{fig:deriv_discontinuity}b, we show this limit and how it is approached as $U$ grows, but
notice also that the Lorentzian shape is approximately correct for all $U$
down to $\Gamma$.  We thus parametrize the XC potential for $U \geq \Gamma$ by the approximate form
\begin{equation}
\tilde{\ve}_{\rm XC} = \alpha\frac{U}{2}\left[1-\langle n_{\rm C} \rangle -
\frac{2}{\pi}\tan^{-1}\left(\frac{1-\langle n_{\rm C} \rangle}
{\sigma}\right) \right],
\label{eq:approx_eps_xc} 
\end{equation}
where $\alpha$ and $\sigma$ are functions of $\Gamma/U$ which $\to 1$ and $8\Gamma/(\pi^2U)$, respectively, in the limit $\Gamma/U \to 0$, and determine the amplitude and width of the correlation contribution
as a function of $\langle n_{\rm C}\rangle$.
Varying $\alpha$ between 0 and 1 corresponds to
 ``turning on'' charge quantization,
 a nonperturbative interaction effect that would not
be described by any local or semilocal approximation.

To determine those parameterizations, we first note the behavior
near 
$\langle n_{\rm C} \rangle=0,\,2$ for $\Gamma \neq 0$.
This is not generic, but  
stems from the restricted Hilbert space of the Anderson model.
For $U/\Gamma\rightarrow \infty$,
Eqs.~(\ref{eq:vs}) and (\ref{eq:KS_n_mol}) imply
that $\partial \ve^{\rm s}/ \partial \langle n_{\rm C}
\rangle \sim \Gamma \pi/(2 \sin^2[\pi \langle n_{\rm C} \rangle/2])$
near 
$\langle n_{\rm C} \rangle=0$.
However, $3\times 10^{-5}[1/\langle n_{\rm C} \rangle^4
]$ fits
the data most accurately over the range $0.1 \leq U/\Gamma \leq 500$ .
Precisely the same form is applied at 
$\langle n_{\rm C} \rangle=2$.
Then
a least-squares fit of Eq.\ (\ref{eq:approx_eps_xc})
to the exact BA results, subtracting the features at 
$\langle n_{\rm C} \rangle=0,2$ and limiting the fit range
to $ 0.1 \leq \langle n_{\rm C} \rangle \leq 1.9$,
yields the values of $\alpha$ and $\sigma$ given in Fig.\ \ref{fig:deriv_discontinuity}c by
the points.

We also found simple fits for these functions, with
\begin{equation}
\begin{split}
\alpha & = \frac{U}{U + 5.68\, \Gamma}, \\
\sigma & = 0.811 \frac{\Gamma}{U} - 0.390 \frac{\Gamma^2}{U^2} - 0.168 \frac{\Gamma^3}{U^3}.
\end{split}
\label{eq:s_ansatz}
\end{equation}
as shown by the smooth curves in Fig.\ \ref{fig:deriv_discontinuity}c.  The corresponding
derivatives are given by dashed curves in Fig.\ \ref{fig:deriv_discontinuity}d. 
Eqs.~(\ref{eq:approx_eps_xc})--(\ref{eq:s_ansatz})
define an interpolation formula for the XC potential which
yields the exact KS potential
in both the 
$U/\Gamma\rightarrow0$ 
and $U/\Gamma\rightarrow \infty$ limits 
and is accurate for all intermediate values of $U \gtrsim \Gamma$,
as shown in Fig.~\ref{fig:deriv_discontinuity}.  
To check that our parametrization is sufficiently accurate, we performed
self-consistent calculations using Eqs.~(\ref{eq:approx_eps_xc}) and (\ref{eq:s_ansatz}), finding transmissions
indistinguishable from the exact ones for all values of $U/\Gamma$ (see Fig.~\ref{fig:occupany}).  
Our interpolation formula shows that the cross-over
between weak and strong correlation occurs for $U \approx 6 \Gamma$.

Our results should prove useful for the development of density
functional theory in general, and for its application to transport through
molecular
junctions.  While much is 
known of the consequences of
the derivative discontinuity in the extreme limit of weak coupling,
our results describe the entire range from strong to weak coupling,
i.e., from weak to strong correlation.  For transport through molecular
junctions, our results provide an exact limit for which both many-body
and DFT approximations can be tested.  
Nor are these just theorists' games with toy models.
For example, for the archetypal Au-[1,4]benzenedithiol single-molecule 
junction, $U > 8\Gamma$.  
In such a system, any approximate XC functional which fails
to account for
derivative discontinuity effects is unlikely to yield
accurate results. 

Note added in proof: After submission of our manuscript, Refs.~\citenum{Evers11b,Evers11a,Stefanucci11} (chronological order) reporting related results appeared on arXiv.  
%
%
This work was supported by DOE under grant number DE-FG02-08ER46496.
KB thanks David Langreth for a decade-long conversation on the subject,
and a lifetime of inspiration.  

\bibliography{refs}

\end{document}